\documentclass[aps, prd, showpacs, floatfix, superscriptaddress, twocolumn, nofootinbib, reprint, longbibliography]{revtex4-2}
\usepackage{multirow, amsmath, hyperref, graphicx, booktabs}
\usepackage[T1]{fontenc}
\usepackage[utf8]{inputenc}
\usepackage{lmodern}
\usepackage{amsfonts}
\usepackage{array}
\usepackage{amssymb}
\usepackage{url}
\usepackage[normalem]{ulem}
\usepackage{mathtools}
\usepackage{textcomp}
\usepackage{parskip}
\usepackage[dvipsnames]{xcolor}
\usepackage[autostyle]{csquotes}
\usepackage{orcidlink}
\usepackage[final]{microtype}
\usepackage{subcaption}
\hypersetup{
colorlinks=true,
citecolor=cyan,
linkcolor=magenta,
urlcolor=NavyBlue,
allbordercolors={0 0 0},
pdfborderstyle={/S/U/W 1}
}
\newcounter{mnotecount}[section]
\newcommand{\mnotex}[1]
{\protect{\stepcounter{mnotecount}}$^{\mbox{\footnotesize
$
\bullet$\themnotecount}}$ \marginpar{
\raggedright\tiny\em
$\!\!\!\!\!\!\,\bullet$\themnotecount: #1} }

\newcommand{\IACS}{School of Physical Sciences, Indian Association for the Cultivation of Science, Kolkata 700032, India}
\newcommand{\IITGn}{Indian Institute of Technology, Gandhinagar, Gujarat 382055, India}
\newcommand{\CEICO}{CEICO, Institute of Physics of the Czech Academy of Sciences, Na Slovance 1999/2, 182 00, Prague 8, Czech Republic}

\newcommand{\CMI}{Chennai Mathematical Institute, Plot H1 SIPCOT IT Park Siruseri, Chennai 603103, Tamil Nadu, India}
\begin{document}

\title{Effect of ultralight dark matter on compact binary mergers}

\author{Kabir Chakravarti}
\email[]{chakravarti@fzu.cz}
\affiliation{\CMI}
\affiliation{\CEICO} 
\author{Soham Acharya}
\email[]{acharyasoham@iitgn.ac.in}
\affiliation{\IITGn} 
\author{Sumanta Chakraborty}
\email[]{tpsc@iacs.res.in}
\affiliation{\IACS} 
\author{Sudipta Sarkar}
\email[]{sudiptas@iitgn.ac.in}
\affiliation{\IITGn} 
\date{\today}

\begin{abstract}
    The growing catalogue of gravitational wave events enables a statistical analysis of compact binary mergers, typically quantified by the binary merger rate density. This quantity can be influenced by ambient factors, following which, in this work we have investigated the impact of dark matter environment on the merger statistics. We construct a baseline astrophysical model of eccentric and quasi-circular compact binary mergers and extend it by incorporating a model of ultra light dark matter, which affects the orbital evolution of binaries through accretion and dynamical friction. Our analysis of the merged population of binary progenitors demonstrates that, compared to the baseline model, ULDM can potentially alter the merger statistics {for ambient non-zero densities, with the best agreement with GWTC-3 obtained for $\rho_0\sim10^6$ GeV/cm$^3$ and eccentricity $\sim0.5$}. A comparison with the gravitational wave data from the GWTC-3 catalogue provides crucial insight into potential observational signatures of the ULDM in merger events, leading to possible constraints on the existence and density of dark matter distribution in galaxies.
\end{abstract}
\maketitle
\section{Introduction}\label{Sec:intro}

Understanding the nature of the compact objects, e.g., black holes (BHs) \cite{Berti:2015itd,Bambi:2019xzp}, neutron stars (NSs) \cite{Vidana:2018lqp,LIGOScientific:2017vwq} and exotic compact objects (ECOs) \cite{Cardoso:2019rvt}, that exist in our universe and assessing their environments has been one of the key research areas of gravitational physics. This will not only provide information about the properties of these compact objects and strong gravity regime that these objects live in, but also about their environments \cite{Barausse_2015,Cardoso:2019rvt}. The upshot being, we will be able to confidently comment, whether general relativity is the theory of gravity describing our universe even in the strong field regime, as well as prove the existence of objects, that are more compact than NSs, but less compact than BHs \cite{Dreyer:2003bv,Psaltis:2008bb,Berti:2018vdi,Cardoso:2016rao}. These will revolutionize the current understanding of our universe in many possible ways . 

However, the significance of studying compact object environments extends beyond these aspects. It can also reveal crucial information about a variety of astrophysical phenomena, including the presence of a third body—such as exoplanets or stars—orbiting these compact objects \cite{10.1093/mnras/254.1.19P, Posselt2009}. Additionally, such investigations can shed light on the existence and distribution of dark matter (DM) \cite{MafaTakisa:2020avv,Cole:2022yzw,Bertone:2024rxe,Cardoso:2021wlq,Chakraborty:2024gcr,Biswas:2023ofz}. In this work, we focus specifically on exploring the presence and properties of DM based on current observational data.


As far as assessing the physics on and around compact objects\footnote{Throughout this work, by compact objects, we will mean objects, whose compactness is larger or equivalent to that of a typical NS.} are considered, the black hole shadow \cite{Perlick_2022} and the gravitational wave (GW) observations  are most effective. While, observations involving black hole shadow are severely limited, in the sense that only two BHs, the M87* and the Sgr A* has been pictured so far \cite{EventHorizonTelescope:2019ths, EventHorizonTelescope:2022urf}, numerous GWs from compact binary coalescence have already been detected through the LIGO-VIRGO-KAGRA (LVK) collaboration \cite{LIGOScientific:2016vbw, LIGOScientific:2017vwq} and the detection rate is expected to grow exponentially with future generations of GW detectors \cite{ET:2019dnz, Reitze:2019iox}. Thus, we will focus on understanding the properties of DM distribution around compact objects through the GW observations.

The fact that our universe entails more invisible matter than visible ones had its origin in the observations involving rotation curves of galaxies, where it was observed that the almost-constant rotation curves can be beautifully explained by postulating the existence of such invisible or, dark matter \cite{Sands:2024vvd}. Besides, several other clinching astrophysical evidences, e.g., bullet cluster, dwarf spheroidal galaxies, gravitational lensing have been put forward in support of the DM hypothesis \cite{Clowe:2006eq, 1987nngp.proc..317F, Massey:2010hh}. Similarly, in the context of cosmology as well, the matter content of our universe requires the existence of DM, whose abundance is at least five times larger than those in the standard model of particle physics \cite{Gaillard:1998ui}. In particular, according to the current paradigm of cosmology, 
DM plays a crucial role in the formation of structures in our universe, as 
they are the first one to collapse into structures called DM halos, into which visible matter falls, leading to individual galaxies. Observational constraints from structure formation \cite{Blumenthal:1984bp}, Lyman-$\alpha$ measurements \cite{Hernquist:1995uma} and lensing \cite{Li:2015xpc} restrict the composition of DM at length scales $\gtrsim100$ Mpc to be cold (CDM). In other words, DM behaves as a pressure-less fluid at these scales. This gives rise to the well-known $\Lambda$CDM paradigm, the standard model of cosmology. However, at small scales, the CDM hypothesis faces serious challenges, the `Core-Cusp problem' \cite{Gentile:2004tb}, the `Galactic Morphology problem' \cite{2010ApJ...723...54K}, and the `Missing satellites problem' \cite{Klypin:1999uc} are to name a few of them. Side by side, the observational constraints on the nature of DM are also much weaker at galactic length scales of $\sim10-100$ Kpc. 

One way to circumvent this difficulty, is to assume that DM form a condensate. This is where, the models of Ultra-Light DM (ULDM) play a pivotal role \cite{Ferreira:2020fam} and can alleviate some of the challenges faced by the CDM scenario. In this formulation, DM is thought to be composed of light bosons, which forms a condensate at sub-halo length scales, while smoothly joins with CDM at larger distances, without violating any observational constraints \cite{Ferreira:2020fam}. Moreover, such light bosons have also been shown to arise from simple extensions of the standard model for particle physics, e.g., models involving axions \cite{Freitas:2021cfi}. All of these make ULDM models a viable extension of the CDM at the galactic scale. 

At the same time, as elaborated above, mergers of compact objects have now become a regular occurrence since their discovery in 2015 by the LVK detectors, which have now detected $\mathcal{O}(100)$ merger events. This presents an interesting possibility to check the existence of ULDM at galactic scales through GW observations. The basics are as follows, in the presence of an environment consisting of ULDM, the binary objects are moving around each other not in vacuum, but in a medium composed of ULDM. Consequently, both the objects in the binary will experience a viscous drag due to the presence of ULDM medium, which can also accrete into the individual objects. The calculation of the viscous drag force in the relativistic context was shown in \cite{Vicente:2022ivh} to be a straight generalisation of the Newtonian dynamical friction formula derived by Chandrasekhar \cite{1943ApJ....97..255C}. It turns out that both the dynamical friction, as well as accretion, as described above, alters the binary inspiral dynamics and leave their signature on the binary phasing. This concept was used \cite{CanevaSantoro:2023aol} to constrain the ambient density of baryons from specific merger events in GWTC2 \cite{LIGOScientific:2020ibl}. More recently another work \cite{Mitra:2023sny} had also used ULDM interactions to distinguish different surface properties of the compact objects, though using extreme-mass-ratio inspirals.

In this work, we start with a simple implication of the previous literature, namely that an environment is likely to alter the merger timescales for a binary system. Given this alteration, we investigate if the ULDM environment can leave its biasing upon the statistics of compact binary mergers. In other words, we want to understand how big or small the effect of ULDM dynamical friction and/or accretion is upon the number of mergers happening per unit redshift per unit observer time per unit co-moving volume, the so-called `astrophysical merger rate density'. Due to the inherent astrophysical uncertainty in the star formation rate, and its modelling, the results presented here will not be conclusive, rather will provide a proof-of-principle that DMs and their properties can be understood using merger rate density. Further astrophysical refinements will make our conclusions more robust, which we will come back to in a subsequent publication. Our paper is organised as follows: in Section~\ref{sec:merg_med}, we review some of the key previous results which are necessary for our subsequent calculations involving the biasing effects expected for mergers in an ULDM environment in Section~\ref{sec:merg_stat}. We take help of toy problems to demonstrate the calculations for individual binaries in Section~\ref{ssec:indbin} and populations Section~\ref{ssec:evpop_prog}. We present our results for the astrophysical merger rate of compact objects in presence of an ULDM environment in Section~\ref{sec:astromerg}. We finally present our conclusions in Section~\ref{Sec:conc}.

\textit{Conventions} --- Throughout this paper we will use geometric units, namely $G=c=1$. Note that we \textit{do not} use $\hbar=1$. Also, we work in four spacetime dimensions, with mostly positive signature convention, such that the flat metric becomes $\eta_{\mu \nu}=\textrm{diag.}(-1,+1,+1,+1)$.

\section{Binary mergers in an ultra-light dark matter environment}\label{sec:merg_med}

In this section, we will present a brief review of the framework and the implications of the ULDM environment on the merger rate of binary compact objects. We start by reviewing the computation of accretion and dynamical friction experienced by a compact object in this environment, following \cite{Vicente:2022ivh, Mitra:2023sny}. Our implementation of these effects into the dynamics of binary mergers and their rate will follow from analogous works involving visible matter environments \cite{CanevaSantoro:2023aol,Roy:2024rhe}.

\subsection{Framework}\label{ssec:frawrk}

We consider the ULDM to be represented by a light scalar field, which satisfies the massive Klein-Gordon equation
\begin{equation}\label{eq:KG}
\Box\Phi - \mu^2\Phi = 0\,,
\end{equation} 
where $m_\Phi = \hbar\mu$ is the mass of the scalar field. In the following analysis we consider $m_\phi = 5 \times 10^{-14}$ eV.  The box operator in Eq.~\eqref{eq:KG} is taken to be a non-trivial background given by the geometry of the massive compact object. If the compact object is non-rotating, the geometry will be that of Schwarzschild, while for rotating compact object the geometry will be given by Kerr metric. Then, following \cite{Mitra:2023sny}, we look for solutions for the scalar field at the rest frame of the compact object, which at large distance from the compact object, behave as plane waves, 
\begin{equation}\label{eq:far_phi}
\Phi(t,r) \sim \sqrt{\frac{\rho_{0}}{\mu}}\,e^{i(\omega t - k_\infty r)}\,. 
\end{equation}
Here, $\rho_{0}$ is the homogenous rest mass density of the scalar field, and $k_{\infty}=\sqrt{\omega^{2}-\mu^{2}}$. In general, the density $\rho_{0}$ of the ULDM, appearing above, will be dependent on the distance from the galactic halo, however, we will assume it to be homogeneous, since our interest is to observe if there is at all any effect of DM on the binary merger rate. Subsequently, one can refine the analysis further, by introducing more realistic density profiles for ULDM. Note that for $\omega>\mu$, the scalar particle scatters off the compact object, while for $\omega<\mu$, the scalar particles form quasi-bound states. In this computation, we will assume that $\omega>\mu$, along with $M\omega\ll 1$, which is the prime reason for studying the ultra-light sector, so that the above conditions are trivially satisfied. The relative velocity of the BH with respect to the DM particles is given by $v^{i}=-k^{i}_{\infty}/\omega$, with $|k^{i}_{\infty}|=k_{\infty}$. Therefore, it follows that, $\omega=\gamma \mu$ and $k^{i}_{\infty}=-\gamma \mu v^{i}$, with $\gamma=1/\sqrt{1-v^{2}}$ is the Lorentz factor. 

The computation of the rate of accretion and dynamical friction follows the below steps --- (a) one solves for the radial part of Eq.~\eqref{eq:KG} in the Kerr background, in the asymptotic and the near-horizon regime; (b) for the near-horizon solution, one imposes reflecting boundary conditions, when these binary objects are stars, and impose purely ingoing boundary conditions for BHs; (c) the asymptotic limit of the near-horizon solution is matched with the near-zone limit of the asymptotic solution; (d) this matching provides ratio of reflected and incident amplitudes of the scalar field, which in turn provides the accretion rate and force due to dynamical friction. In the frame of the object moving through DM distribution, these read \cite{Mitra:2023sny,Vicente:2022ivh},
\begin{align}
\dot{M}&=\int_{\infty}r^{2}T_{tr}d\Omega_{2}=\rho A_{\rm h}\left(\frac{e^{\pi \eta}\pi\eta}{\sinh(\pi \eta)}\right)\left(\frac{1-R}{1+R}\right)~,
\\
F^{i}&=\int_{\infty}r^{2}T^{i}_{r}d\Omega_{2}=-\frac{4\pi \rho M^{2}}{v^{2}}(1+v^{2})^{2}\mathcal{D}\zeta^{i}-\dot{M}v^{i}~.
\end{align}
The quantity $R$, appearing above, in the expression for $\dot{M}$, is unity for stars, and vanishes for BHs. Thus the above expression for accretion is relevant for BHs alone. Here, $T_{\mu \nu}$ is the energy-momentum tensor of the scalar field, $\zeta^{i}$ is the direction from which the scalar waves impinges on the BH at rest, $\eta=-\{\gamma \mu M(1+v^{2})/v\}$, $\rho=\gamma^{2}\rho_{0}$ and $\mathcal{D}=\ln \Lambda-\textrm{Re}[\Psi(1-i\eta)]$. The quantity $\Psi(x)$ is the di-gamma function and $\Lambda=\gamma v\mu b_{\rm max}$, with $b_{\rm max}$ being the maximum impact parameter. Note that $\gamma v\mu$ is the inverse of the de-Broglie wavelength associated with the DM particles. Finally, the quantity $A_{\rm h}=4\pi(r_{+}^{2}+\mathrm{a}^{2})$ is the horizon area, with $r_{+}=M+\sqrt{M^{2}-\mathrm{a}^{2}}$, being the location of the event horizon, and $J=\mathrm{a}M$ is the angular momentum of the BH. In the expression for force experienced by the compact object due to scattering from ULDM, the first term represents the gravitational drag from the surrounding ULDM field, and the second term accounts for momentum loss due to accretion.

The above analysis was performed in the frame of the compact objects, while the observation is being made at asymptotic infinity. Since the DM is static in the observer's frame of reference, we need to make a transformation from the frame of the BH to the frame of the observer. This is simply achieved by the following transformations: $\dot{M}'=(\dot{M}/\gamma)$ and $F'_{i}=F_{i}+\dot{M}v_{i}$. Then, it turns out (see \cite{Mitra:2023sny} for the detailed calculation) that the accretion rate of an object of mass $M$ and angular momentum $J$ moving through the ULDM medium with an orbital speed $v$, in the rest frame of the DM, is given by,
\begin{equation}\label{eq:m_dot}
\dot{M}^{\prime}=\hbar n \omega A_{+}\left(\frac{e^{-\pi \eta} \pi \eta}{\sinh (\pi \eta)}\right) \left(\frac{1-R}{1+R}\right),
\end{equation}
where, $n=(\rho_0/m_\Phi)$ is the number density of the ULDM particles for observers at rest. We assume complete absorption if the object is a BH ($R=0$) and complete transmission in all other cases ($R=1$). Similarly, the dynamical friction force acting on the object in the rest frame of the DM is given by
\begin{align}\label{eq:f_dyn}
\vec{F}^{\prime}&=-\frac{4\pi \rho_0 M^{2}}{v^{2}}(1+v^{2})^{2}\mathcal{D}\vec{\zeta}=-\frac{4\pi\hbar n \eta^{2}}{\mu}\Delta\Psi\vec{\zeta}\,,
\end{align}
where, $\mathcal{D}=\Delta\Psi=\mathrm{Re}[\Psi(1+\ell_m+i\eta)-\Psi(1+i\eta)]$ with $\Psi$ being the digamma function and $\ell_m$ the cut-off in the angular harmonic corresponding to the maximum impact parameter of the object $b_{\rm max}$. As evident, the dynamical friction is independent of the details of the nature of the compact object, while accretion is dependent on the compact object. We will show the effect of accretion on binary merger rate in Section~\ref{ssec:impbin}. Multiplying the dynamical friction force by the relative speed $v$ between the compact objects, we can finally deduce the power dissipation through the dynamical friction as
\begin{equation}\label{eq:pdf}
P_\mathrm{df}=\vec{F}'\cdot\vec{v}=\frac{4\pi\hbar}{\mu}nv\eta^2\Delta\Psi~.
\end{equation}
This dissipative term will add up to the quadrupolar radiation emitted from the binary inspiral, leading to faster energy loss from the system. This results into faster inspiral, and hence reduces the merger time, having impact on the rate of binary merger. We will implement this for (both circular and elliptic) binary orbits in the subsequent section. 

\subsection{Implementation in binary inspiral}\label{ssec:impbin}

Following the outlines of Section~\ref{ssec:frawrk}, we note that the motion of binary objects through the DM environment creates an extra dissipation force, which in the scalar's rest field gives rise to an additional loss of power, given by $P_\mathrm{df}$. In the case of a binary orbit, the consequence being, the loss of the total orbital energy of the binary. In addition to this dissipation due to DM, there is also a dissipation from the emission of GWs, which at leading order is given in the case of quasi-circular orbits by \cite{Will:1996zj},
\begin{equation}\label{eq:pgw}
P_\mathrm{gw}=\frac{32}{5}\frac{m_{\rm r}^{2}M_{\rm T}^{3}}{a^5}\,,
\end{equation}
where, $m_{\rm r}=M_{1}M_{2}/(M_{1}+M_{2})$ and $M_{\rm T}=M_{1}+M_{2}$ are the reduced mass and the total mass of the particular binary, respectively, with individual masses being $M_{1}$ and $M_{2}$. The quantity $a$ is the binary separation and is related to the orbital velocity $v$, as well as the orbital frequency $\Omega$ of the reduced mass system by $v^{2}= (M_{\rm T}/a)=(M_{\rm T}\Omega)^{2/3}$. Note that the frequency of the emitted GW is twice of the orbital frequency. Following the implementation of \cite{Roy:2024rhe}, the equation for quasi-circular orbital phasing of GW becomes
\begin{align}\label{eq:orb}
\frac{d\phi}{dt}&=\Omega\,,
\nonumber 
\\
\frac{dE_{\rm orb}}{dt}&=-P_\mathrm{gw}-P_\mathrm{df}\,,
\end{align}
where, the orbital energy $E_{\rm orb}$ is given at the leading order by $E_{\rm orb}=-(1/2)m_{\rm r}v^{2}$. We can then substitute for $E_{\rm orb}$ along with the expressions of $P_\mathrm{df}$ and $P_\mathrm{gw}$ from Eq.~\eqref{eq:pdf} and Eq.~\eqref{eq:pgw} to obtain, 
\begin{equation}\label{eq:num_evol}
-\frac{1}{2}\frac{d}{dt}\left(\frac{m_{\rm r}M_{\rm T}}{a}\right)=-\frac{32}{5}\frac{m_{\rm r}^{2}M_{\rm T}^{3}}{a^5} - \frac{4\pi\hbar n}{\mu} \sum_{i=1}^{2}\eta^{2}_{i}v_i\Delta\Psi_i\,,
\end{equation}
where the sum over $i$ represents a summation over the components of the binary. Then it is easy to see from the motion of the reduced mass that $v_{i}=(m_{i}/M_{\rm T})v= (m_{i}/M_{\rm T})\sqrt{M_{\rm T}/a}$, where $i,j$ are just distinct labels for the binary components and not spatial indices. We should also note that the time differentiation operator on the left hand of Eq.~\eqref{eq:num_evol} will also hit both $m_{\rm r}$ and $M_{\rm T}$, where the corresponding expressions have to be substituted from Eq.~\eqref{eq:m_dot}. In this way we see that we end up with a differential equation in terms of the binary's orbital separation $a$. We must however notice that Eq.~\eqref{eq:orb} and hence Eq.~\eqref{eq:num_evol} are specific to quasi-circular inspirals only. For binary orbits starting with an ellipticity, the evolution equation becomes a set of two coupled linear differential equations in the orbital averaged separation $\langle a\rangle$ and the eccentricity $e$. In this work, we have used the framework of [cite Peters Matthews] to calculate inspirals for elliptic cases. The necessary computations have been shown in the Appendix~\ref{app:final}. The numerical evolution of Eq.~\eqref{eq:num_evol} and its elliptic counterpart Eq.~\eqref{eq:coupled} determine the dynamics of the binary in question, which will be different from the vacuum evolution, due to the presence of dynamical friction.

\section{Dark matter environment and merger statistics}\label{sec:merg_stat}

Here, we will present our results regarding merger statistics of binary evolution in the environment of ULDM and shall quantify the effects of ULDM on compact binary mergers. We start by discussing the effects of environment on individual binaries and then focus on the effect on binary population and their merger rate.

\subsection{Individual Binaries}\label{ssec:indbin}

We show the effect of the presence of ULDM on the merger timescale of individual binaries in this section. For that we start with a chosen mass configuration of binary main sequence stars as progenitors and track their merger status by evolving Eq.~\eqref{eq:orb} to obtain the binary separation $a$ with time. To do this consistently, we require a proper transformation function when a progenitor star reaches its stellar endpoint. We implement this by a simple mass-transfer function \cite{2012ApJ...749...91F} to determine the mass of the evolutionary remnant of the main sequence star. In reality, such endpoints of massive main-sequence stars are usually accompanied by novas, which are highly model dependent and is likely to add a further layer of uncertainty to our baseline model, but will not significantly contribute to an understanding of the effects of ULDM on binary merger. Nevertheless, we need to implement a model to facilitate mergers and not disrupt them. To do this, we adopt a method similar to the evolution method  previously implemented in \cite{Chakravarti:2023wlc} with the notable exception that our method does not implement a constant ejecta velocity to model the kick imparted to the stellar remnants. Rather, we let the ejecta velocity to depend on the initial masses of the stellar progenitors and also on the associated mass loss. In this case, the kick imparted to the remnant is likely to cause a sudden change to the angular momentum, depending on various parameters of the system. The change in angular momentum also causes a change in the angular velocity of the remnants. In the case of quasi-circular inspirals, these changes can be determined through the following equations obtained from $L\coloneqq m_{\rm r}a^2\Omega$ and Kepler's third law $\Omega^{2}a^{3}=M_{\rm T}$.
\begin{align}\label{eq:omega0}
\left(\frac{\delta \Omega}{\Omega}\right)&=-3\left(\frac{\delta L}{L}\right)
+3\left(\frac{\delta m_{\rm r}}{m_{\rm r}}\right)+2\left(\frac{\delta M_{\rm T}}{M_{\rm T}}\right)\,.
\end{align}
Here $\delta x$ represents the change in a quantity $x$ before and after the stellar endpoint. From the mass-transfer function implementation of \cite{2012ApJ...749...91F} the final mass is less than the initial mass because a nova usually involves ejection of mass. Therefore, it follows that $(\delta M_{\rm T},\delta m_{\rm r})<0$. Using these results in Eq.~\eqref{eq:omega0}, we notice that only if $\delta L<0$, one might have $\delta \Omega>0$. Then to connect with the binary separation at the stellar endpoint, we use Kepler's third law again to obtain,
\begin{align}\label{eq:omega}
3(\delta a/a)=(\delta M_{\rm T}/M_{\rm T})-2(\delta \Omega/\Omega)
\end{align}
This suggests that whenever $(\delta\Omega/\Omega)\geq0$, the binary separation always decreases at stellar endpoint. Therefore we work with precisely those events where $(\delta\Omega/\Omega)\geq0$ is obeyed at mass transfer, which in turn ensure that the binary separation never \textit{increase} across the stellar endpoint. We note that the condition $\delta L <0$ implies that the angular momentum induced by the ejecta is opposite to that of the orbital angular momentum. On the other hand, the condition $\delta L\geq0$ increases the binary separation and can lead to potential disruption of the system \cite{Belczynski:2016obo}. Hence on average, the number of binaries whose separation between the individual components decrease will be reduced by some factor. Luckily, this decrease in the number of such events does not affect our results because we rescale the merger events so that the local merger rate density matches observations (see Sec ~\ref{sec:astromerg}).

Further, from the above expressions it follows that $|\delta a/a|\geq (1/3)|\delta M_{\rm T}/M_{\rm T}|$. From this, we finally make our choice of $\delta a$ across stellar endpoint as
\begin{equation}\label{eq:del_a}
\delta a = \left(\frac{\delta M_{\rm T}}{M_{\rm T}}\right)a\,.
\end{equation}
For the elliptic case, the analogous calculation to compute the change in orbital separation and eccentricity are straightforward but tedious. We have reproduced the necessary calculations in \ref{app:eccentric_endpoint}. It turns out that the elliptical cases have the same $\delta \langle a\rangle$ across the stellar endpoint as in Eq.~\eqref{eq:del_a}, while $\delta e = 0$. Also we note from Eq.~\eqref{eq:del_a} that $\delta M_{\rm T}<0$, implies $\delta a<0$, since both $a$ and $M_{\rm T}$ are positive definite quantities. This completes the formulation of our baseline model, which is constructed in the absence of any ULDM environment. 

Before moving further we note that the overall effect of the ULDM environment is through accretion and dynamical friction. As already emphasized, Eq.~\eqref{eq:f_dyn} depicts that the drag term acts universally upon any object, but the accretion term is dependent upon the nature of the surface reflectivity, which will vanish for a BH, while will be unity for a star. Thus the evolution of the binary separation will be different for stars and BHs. With this specific model, we consider the following example involving a binary progenitor configuration of total mass and mass-ratio being $M_{\rm T}^{\rm P}=45 M_\odot$ and, $q^{\rm P}=0.8$, respectively, for two different starting redshift and for two cases with and without ULDM environment. Quantities related to the progenitor stars will be denoted with a superscript $\textrm{P}$. At this point we remark that further simplicity is maintained in our baseline model by not considering properties like BH spins and/or tidal interactions during matter phases. Under these simplifying assumptions, any difference in the final binary configuration between the baseline model and the model with ULDM can be thought of as being solely due to the presence of the ULDM environment. 

\begin{figure*}[t!]
    \centering
    \includegraphics[scale = 0.5]{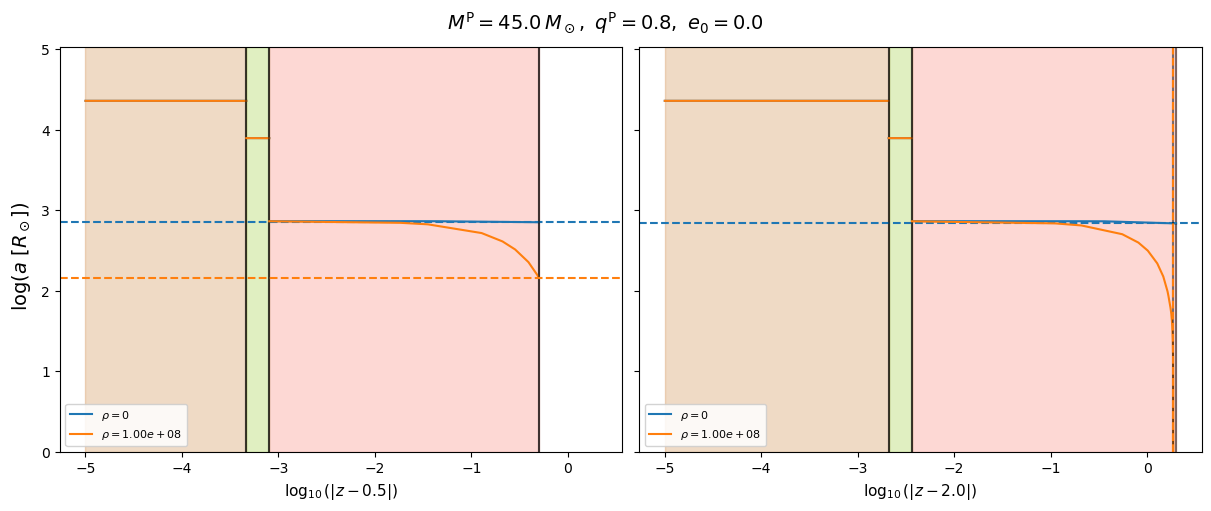}
    \caption{Quasi-circular evolution of the binary separation parameter $a$ (in units of solar radius $R_{\odot}$) of a given progenitor as a function of the redshift $z$ with ($\rho_0 > 0$) and without ($\rho_0 = 0$) the presence of ULDM. The left and right panels show the evolution starting from $z = 0.5$ and $z =2.0$ respectively. The units of DM density $\rho_0$ are in $\textrm{GeV}/\textrm{cm}^{3}$.}
    \label{fig:progmerg0p2}
\end{figure*}

The results for the quasi-circular evolution of the orbital separation of the binary system, with progenitors having $M_{\rm T}^{\rm P}=45 M_\odot$ and, $q^{\rm P}=0.8$, have been presented in Fig.~\ref{fig:progmerg0p2}, starting at redshift $z=0.2$ (left panel) and $z=2.0$ (right panel). The evolutionary phases involve the binary components to be either stars or compact objects, e.g., NSs and BHs. Starting from given redshift on the left hand side of each panel of Fig.~\ref{fig:progmerg0p2}, the different shadings correspond to the binary in its different phases. The binary start from the star-star phase (brown shade), the compact object-star phase (green phase) and finally both components turn into compact objects (the crimson shade). As per our elaboration of the baseline model, we notice two discontinuous jumps in the binary separation $a$ across the two endpoints of the main-sequence stars in the binary, following the implementation in Eq.~\eqref{eq:del_a}. These endpoints of the stars are determined through the mass-luminosity relation of main sequence stars \cite{2005essp.book.....S}, which allows us to compute a simple scaling relation for the lifespan $\tau_{\rm star}$ of a star with its progenitor mass $M^{\rm P}$, namely $\tau_{\rm star}\propto {M^{\rm P}}^{-2.5}$. Dividing the corresponding quantities for the Sun, and keeping in mind that the estimated life span of Sun is $\tau_\odot \approx 10$ Gyr, we obtain the typical lifespan of a star in the binary to be,
\begin{equation}\label{eq:lfspn}
\tau^{i}_{\rm star}(\textrm{in}~\mathrm{Gyr})=\frac{10}{({M^{\rm P}_{i})}^{2.5}}\,.
\end{equation} 
Here, $i$ refers to the binary components. We note that owing to the source on the left starting from $z=0.2$, and the main-sequence stellar lifespan (c.f Eq.~\eqref{eq:lfspn}), the binary remains unmerged at $z=0$ even in the presence of ULDM. However, even the unmerged binaries will carry a signature of the DM environment in their present orbital separation $a_{0}\equiv a(z=0)$, which for the ULDM case is significantly less than its value for the vacuum configuration as depicted by the horizontal dashed lines in the left panel of Fig.~\ref{fig:progmerg0p2}). We also point out that unmerged binaries are effectively binaries in their early inspiral, which cannot be observed by the ground based GW detectors. But, un-merged binaries will show up as nearly monochromatic GWs in the space based detectors, e.g., LISA \cite{Karnesis:2022vdp} and DECIGO \cite{Kawamura:2020pcg}. Given that the orbital separation $\mathrm{a}$ is highly dependent on the desnity of DM particles in the environment, independent constraints on the ULDM can be obtained from the observation of low-frequency GW signals by LISA \cite{Duque:2023seg}.

In the right panel of Fig.~\ref{fig:progmerg0p2}, we show a case where the binary progenitors start at a higher redshift, namely at $z = 2.0$. In contrast to the previous case, we see that the presence of the ULDM compels the binary to merge while the vacuum case does not merge. This behaviour is expected, and in consonance with our intuition from earlier. In short, the dissipation from the ULDM environment adds to the energy loss of the binary, over and above the energy loss from GWs, and consequently we are left with faster merger timescales and/or smaller remnant separation.

\subsection{Evolution of a binary population of progenitors from a given redshift}\label{ssec:evpop_prog}

We have discussed in the previous section how the ULDM environment affects a pair of compact objects undergoing a binary inspiral. In this section we will present the corresponding result for a population of binary progenitors. In doing so, the individual pair of objects can now be thought of as building blocks of a population. To characterise the statistics of merger populations, we need to answer to the question, what happens when many of such individual binaries evolve under a ULDM environment simultaneously. We will answer this question by first considering a toy problem, namely, we will determine the final configuration of a population of binary main-sequence star systems, all of which began at the same redshift. All of these binary systems, during their evolution, undergo accretion from the ULDM environment (if they gets converted to BHs) as well as dynamical friction. In addition to these, they will lose energy by emission of GW. The rate of dissipation of the orbital energy, therefore, depends on all the binary parameters (namely masses, spins and eccentricity) as well as on the environment the binary system lives in. Hence, it is clear that the orbital evolution of the binaries will non-trivially depend on the progenitors mass distribution $\{M_{i}^{\rm P}\}$ (the so-called initial mass function or, IMF), the spin distribution $\{\chi_{i}^{\rm P}\}$ ($\chi\equiv (J/M^{2})$, where $J$ is the angular momentum) and the eccentricity distribution $\{e^{\rm P}\}$ of the population of the progenitor stars and their evolved remnants. The IMF is for the most part unknown, but is expected to be intricately linked to astrophysical factors like stellar metallicity and accretion. The only estimates of the IMF comes from observations of stars in our own galaxy \cite{Kroupa:2000iv} and is usually taken to follow a broken power-law distribution with less massive stars being more probable. Without observation of individual stars, very little can be known about the nature of the IMF \cite{2013pss5.book..115K, 2018A&A...620A..39J} in a model independent way at the redshifts of star formation $z\gtrsim1$. Additionally we also see from the GWTC-3 \cite{KAGRA:2021vkt} catalogue that the spin distributions of the remnants cannot still be accurately inferred, while the distributions about eccentricity remain unknown. Faced with this uncertainty, we will opt for the most simplistic or minimalistic formulation possible with respect to a given set of circumstances. 

Consequently, we will work with both uniform as well as non-unform (Kroupa) IMFs \cite{Kroupa:2002ky} in our analysis. The uniform IMF, though probably unphysical still allows us to be  maximally agnostic, while the Kroupa model is motivated by the stellar observations from the galactic plane. Then, for the eccentricity we model its effects by evolving eccentric binary main-sequence progenitor stars. Since the distribution of eccentricity is unknown, we consider a number of eccentricities while assuming the same starting eccentricity for all progenitor binaries. Also, we will continue to assume no effects of BH spin in our analysis. Moreover, our evolution does not include other astrophysical effects, like Roche overflows, stellar winds, common envelope binary evolution or the peculiar motion associated with the centre of mass of the binaries. We remark that factoring in any (or all) of these components require a fully numerical population synthesis involving hydrodynamical and/or N-body simulations, which is beyond the scope of this work. Therefore, what we are assuming is that the dynamical friction and energy loss by GWs are the only dissipative components in the dynamics. So, any constraint on the density of the ULDM from our analysis should be treated only as an upper limit.

Let us now focus on our choice of lower and upper cut-off for the IMFs. It is reasonable to expect that progenitor stars which have masses below a limit cannot produce compact remnants. Keeping this in mind we choose a lower cut-off of the progenitor masses given by $M_i^{\rm P}\geq 10 M_\odot$. Then the use of the uniform IMF requires us to also choose an appropriate upper bound in the progenitor stellar-mass distribution. Masses of stars are bounded from above to $M_{i}^{\rm P}\leq 100 M_\odot$ from considerations of hydrostatic equilibrium. For our purposes in this work, we choose a slightly lower value for the upper limit of the progenitor masses, namely, $M_{i}^{\rm P}\leq 50 M_\odot$. A limitation of this mass cut-off could be that it might produce bias for the choice of uniform IMF. However, the bias for the case of the more realistic non-uniform IMF will be negligible, as that IMF function decays to zero at those high mass values. 

\begin{figure*}[ht!]
    \centering
    \includegraphics[width=\textwidth]{Final_Plots/Phase_Plots.pdf}
    \caption{Plot showing the composition of merged and unmerged final states (phase plot) of all possible combinations of binary main sequence star progenitors beginning at $z=0.5$ for circular (left) and elliptic cases. The grey shaded regions in each individual  phase plot  are off-limits in progenitor configuration in our modelling. We show the $z$ distribution for the merged cases (green shades) and the binary separation $a$ in units of solar radius $\mathrm{R}_\odot$ at $z=0$ for the unmerged cases as a function of the total progenitor masses in the binary and assuming a mass transfer, for different cases of dark matter density. The red line is the separatrix between the unmerged and merged populations at $z=0$. See text for further details.}
    \label{fig:progmerg2}
\end{figure*}

A final uncertainty which plays a role in determining the merger timescales is the choice of initial separation of the progenitor stars. However, there are still some physically motivated factors which help to narrow down the range of the initial separations. We note that the initial separation is bound from below if we demand no Roche overflow in the initial progenitor configuration. Consequently we assume the initial configuration of the binaries to always obey the no Roche overflow condition. We have implemented this by making the sum of the initial Roche lobes \footnote{For elliptical orbits the Roche radius is computed with respect to the semi-major axis of the initial configuration} of the component stars $1.5$ times the sum of their radii. The rescaling of merger rate (see Sec ~\ref{sec:astromerg}) ensures that the rate of mergers in our model is commensurate with observations even with our specific choice of initial separations. We now return to the question of the final state of main sequence star progenitors, all of which evolve in the presence of ULDM, while starting at a fixed redshift. We present the answer to this question in Fig.~\ref{fig:progmerg2}, where we show that the end states of the evolution for progenitor binaries starting from $z=0.5$. Here and henceforth, we adopt the convention that for the progenitors $M_2^{\rm P}\leq M_1^{\rm P}$, such that the mass-ratio $q^{\rm P}\leq 1$. Following which, in Fig.~\ref{fig:progmerg2}, we show four different compositions of merged and unmerged final states (phase plots) of binary evolution corresponding to four different ambient densities of the ULDM environment for quasi-circular (left) and elliptic cases. For each of the phase plots, we show the mass ratio of the progenitors $q^{\rm P}$ on the x-axis and the total mass of the binary configuration $M_{\rm T}^{\rm P}=M_{1}^{\rm P}+M_{2}^{\rm P}$ on the y-axis. Following our reasoning above, the grey shaded regions represent configurations that are off-limits to our interest, i.e., with progenitors masses being too high, or, too low. It is expected that the merger timescales for the binary systems will not all be the same, and thus we would get a combination of merged and unmerged binary populations, depending on the DM density in the environment. In every phase plot of Fig.~\ref{fig:progmerg2}, the red line acts as a separatrix, which separates the configuration space into merged and unmerged endpoints for the binaries. The redshifts for the merged population are shown by the green-yellow shade, while for the unmerged populations we depict the binary separation at $z=0$ by the blue shade. 

As evident from Fig.~\ref{fig:progmerg2}, an increase in the ULDM density \footnote{We have shown circular and elliptic cases with different orders of ULDM density in order for the phase plots to be informative across the range of variation of ULDM density} leads to a reduction in the area under the blue-shaded region for both circular and eccentric cases, which, as expected, implies more mergers occurring at higher redshifts. Additionally, we observe that a higher ULDM density facilitates the merging of remnants that would otherwise remain unmerged, particularly those originating from more massive progenitors. The oscillatory patterns seen in the parameter space of the unmerged population for $\rho_0\geq 7.5\times10^7~\textrm{GeV}/\textrm{cm}^{3}$ for the circular case (and for $\rho_0\geq 5\times10^4~\textrm{GeV}/\textrm{cm}^{3}$ for eccentric case) arises from similar oscillatory non-linearities present in our assumed mass-transfer function. 

\begin{figure}
    \includegraphics[width=\columnwidth]{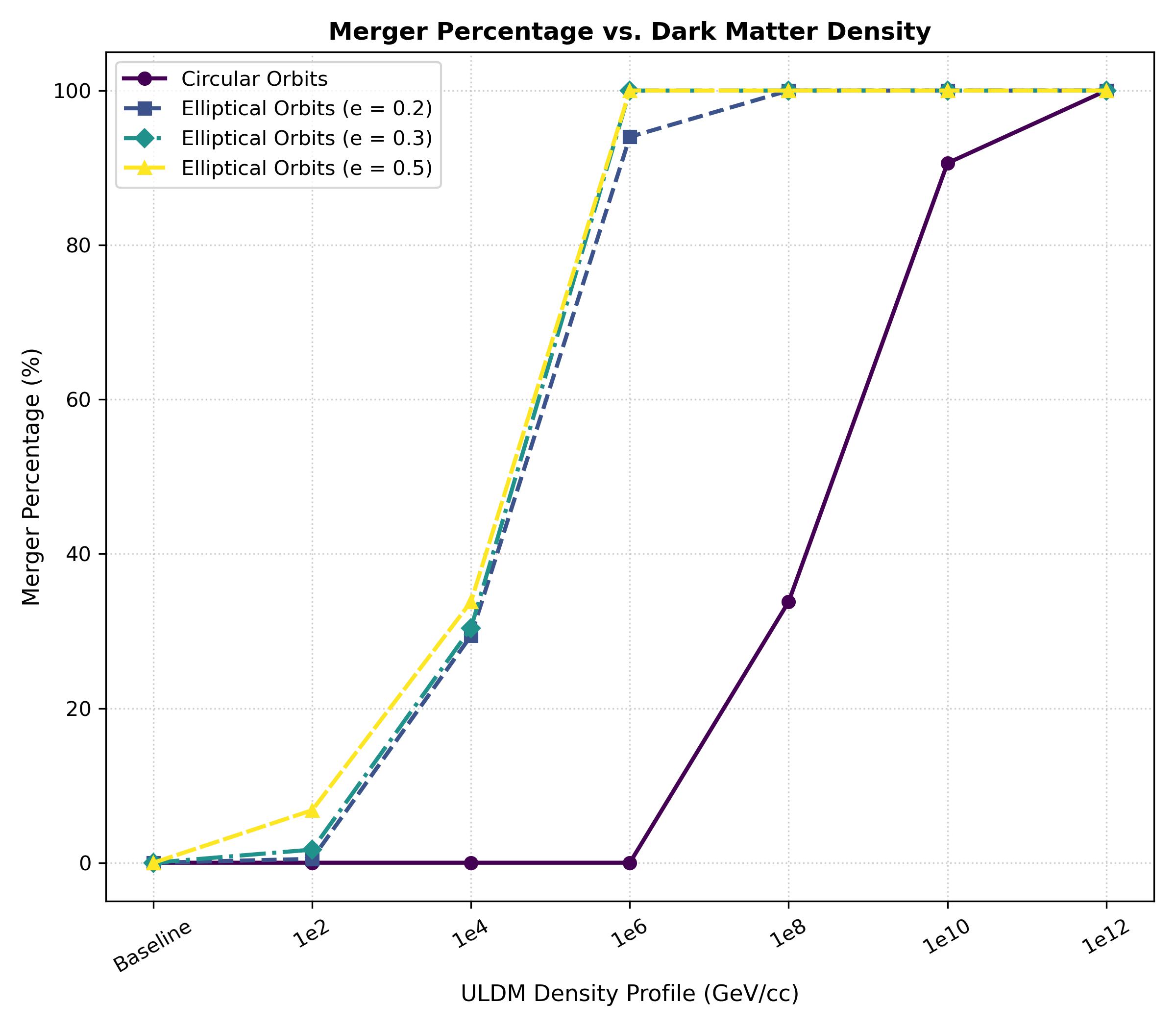}
    \caption{Plot showing the fraction of merged binaries as a function of ambient ULDM density for eccentric and quasi-circular cases for progenitors starting at $z=0.5$. Units of DM density $\rho_0$ are in $\mathrm{GeV}/\textrm{cm}^{3}$}.
    \label{fig:progmerg_ratio}
\end{figure}

To further illustrate these effects, we have plotted the fraction of the merged configurations along with the ULDM densities within which they evolve in Fig.~\ref{fig:progmerg_ratio}. From this plot we can clearly identify that ULDM densities approaching $\rho_0\sim 10^{12}\mathrm{GeV}/\textrm{cm}^{3}$ causes all available binaries to merge, while $\rho_0\lesssim 10^{2}\mathrm{GeV}/\textrm{cm}^{3}$ are too rare to significantly alter the merger fraction. Moreover we also see that given a ULDM density more elliptic binaries merge more. For densities $\rho_0\gtrsim 5\times10^8 \mathrm{GeV}/\textrm{cm}^{3}$, ULDM begins to have a significant impact on the merger fraction in the quasi-circular case leading to nearly $\gtrsim60\%$ of configurations merging. Whereas the same number is attained for much smaller densities $\sim5\times 10^{4}\mathrm{GeV}/\textrm{cm}^{3}$ for the eccentric cases. These findings indicate that, within the framework of our simplified toy model and irrespective of eccentricity, the presence of ULDM noticeably influences the evolution of the compact binary populations, systematically increasing the fraction of mergers occurring at earlier cosmic times. While this model is intentionally minimalistic, it serves as an illustrative tool to study how the surrounding dark matter environment can shape merger statistics.

\section{Effects on Astrophysical Merger Rates}\label{sec:astromerg}

The results of the previous section have firmly established that presence of an ULDM environment should have an effect on the population statistics of mergers. However, we only considered progenitor sources at a single redshift $z = 0.5$. A complete picture can be obtained if we can repeat the above exercise, but factoring in sources from a given range of redshifts. Additionally, we recall that all redshifts will not contribute equally to the generation of compact binary progenitors. This is because the comoving density of main-sequence stars is not constant with the redshift, and has a functional dependence which is well-known in the literature as the Star Formation Rate (SFR). In our analysis in this section, we have used the SFR function, as proposed in Madau \& Dickinson \cite{Madau:2014bja}. At this outset, we would like to emphasize that the SFR simply provides the total mass of the stars produced at a given redshift per comoving volume. One needs to convolve it with the IMF in order to get the number density of the stars. For the cases we have considered, the convolution integrals will be non-trivial for the Kroupa IMF and just a constant factor in the case of flat IMF. Subsequent multiplication with the comoving volume results into a function for the number of sources in the redshift bin $z$ and $z+dz$, which can be observed per unit observer's time. This function, representing number of binary progenitors per unit redshift per unit time, is given by
\begin{equation}\label{eq:source_rate}
\frac{d^2N}{dzdt}=\frac{1}{1+z}\left(\frac{dV_{\rm c}}{dz}\right)\mathcal{R_\psi}(z),  
\end{equation} 
where $\mathcal{R_\psi}(z)$ is the SFR at redshift $z$, and $(dV_{\rm c}/dz)$ is arising from the cosmological comoving volume. The factor $(1+z)^{-1}$ appears because of the difference in the time intervals at the source and at the observer. Given Eq.~\eqref{eq:source_rate}, we are now in a position to compute the rate of binary compact mergers for our underlying astrophysical model. To compute the `merger-rate density' we need to first evolve all sources at a given $z$, given by Eq.~\eqref{eq:source_rate}, either to their merger or to $z=0$. The evolution has to be done for both the eccentric and the quasi-circular cases. We then bin the population of all the merged events according to their merger redshifts, which gives for us the number $(d^2N_{\rm m}/dz_{\rm m}dt)$, depicting the merger events between redshifts $z_{\rm m}$ and $z_{\rm m}+dz_{\rm m}$, observed per unit observer's time. Here, $z_m$ represents the merger redshifts. Dividing the above by the comoving volume factor at $z_m$ and multiplying by a factor of $1+z_{\rm m}$ to account for the redshift factor, we finally arrive at our estimate for the merger rate density,
\begin{equation}\label{eq:rm}
\mathcal{R}_{\rm m}(z_{\rm m}) \sim \frac{d^2N_{\rm m}}{dz_{\rm m}dt}\left(\frac{dz_{\rm m}}{dV_{\rm c}}\right)\left(1+z_{\rm m}\right)~. 
\end{equation}
We can now compare our results and match our estimates of the merger rate $\mathcal{R}_{\rm m}(z_{\rm m})$ to the inferred rate $\mathcal{R}_{\rm m}^\mathrm{inf}(z_{\rm m})$ from the LIGO-Virgo observations. But before we do that a word of caution is in order. We reiterate that our baseline model does not consider inputs from many dissipative astrophysical effects (like Roche overflows, common envelope evolutions etc.). It also does not contain information about other non-dissipative astrophysics like for example the $z$ dependence of the stellar IMF or the fraction of progenitors that get trapped in binary systems. Therefore, it is natural to expect an incorrect estimation for our calculated merger rates. We should also remark that at current sensitivity, GW observations can only observe the `local' merger rate density $\mathcal{R}_{\rm m}^\mathrm{obs}(0)$, i.e., the merger rate density at $z=0$. The inaccuracy in the estimation of the merger rate can nevertheless be removed by an overall rescaling of the number of sources such that the rescaled estimates of the rate $\mathcal{R}_{\rm m}(z_{\rm m})$ obey observational bounds at $z_{\rm m} = 0$. In other words, such a rescaling will make $\mathcal{R}_{\rm m}(0)$ lie within the error bars of $\mathcal{R}_{\rm m}^\mathrm{obs}(0)\equiv R_{0}$. Subsequently, following this standard procedure in the population synthesis literature (see, for example \cite{Calore:2020bpd}) we make use of appropriate rescaling functions to bring the value of $\mathcal{R}_{\rm m}(0)$ within the GWTC-3 bounds \cite{KAGRA:2021duu}. 

\begin{figure*}[!t]
    \centering
    \includegraphics[width=0.8\textwidth]{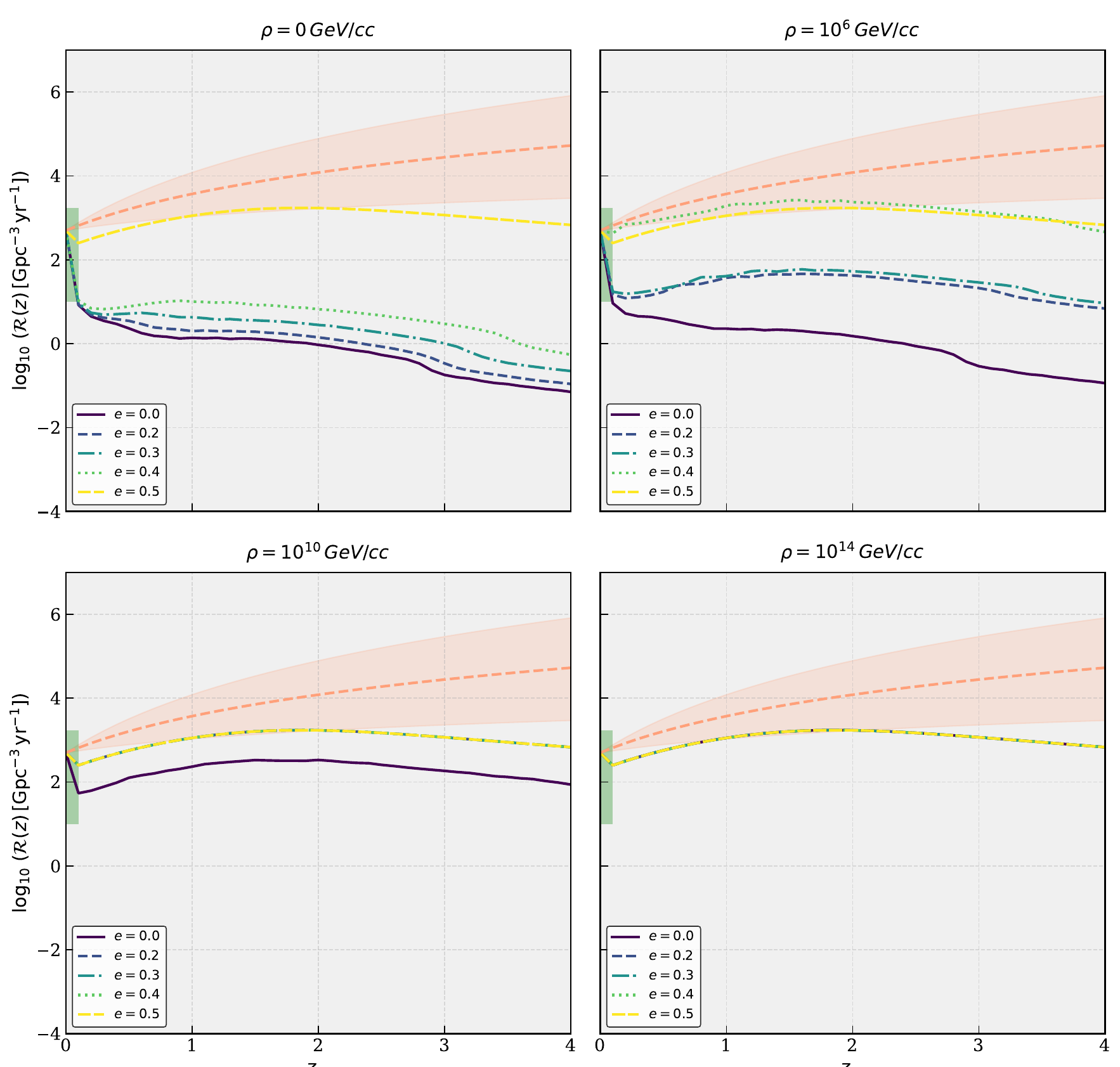}
    \caption{Plot showing the merger rate densities $\mathcal{R}_m\left(z_m\right)$ as a function of redshift $z_m$ for our choices of ULDM density $\rho$ and initial eccentricity $e_0$. The different colored curves correspond to different values of $e_0$, as indicated in the legends. The green shaded region represents our current uncertainty in the total local merger rate density $R_0$ of binary compact objects inferred from GWTC-3 data \cite{LIGOScientific:2020ibl}. The orange dashed line shows the inferred redshift evolution $\mathcal{R}_m^{\mathrm{inf}}\left(z_m\right)$ from GWTC-3 data, with the orange shaded region representing the corresponding uncertainty. See text for more details and \ref{fig:kroupa_imf} for merger rate density corresponding to Kroupa IMF.}
    \label{fig:flat_imf}
\end{figure*}

\begin{figure*}[!t]
    \centering
    \includegraphics[width=0.8
    \textwidth]{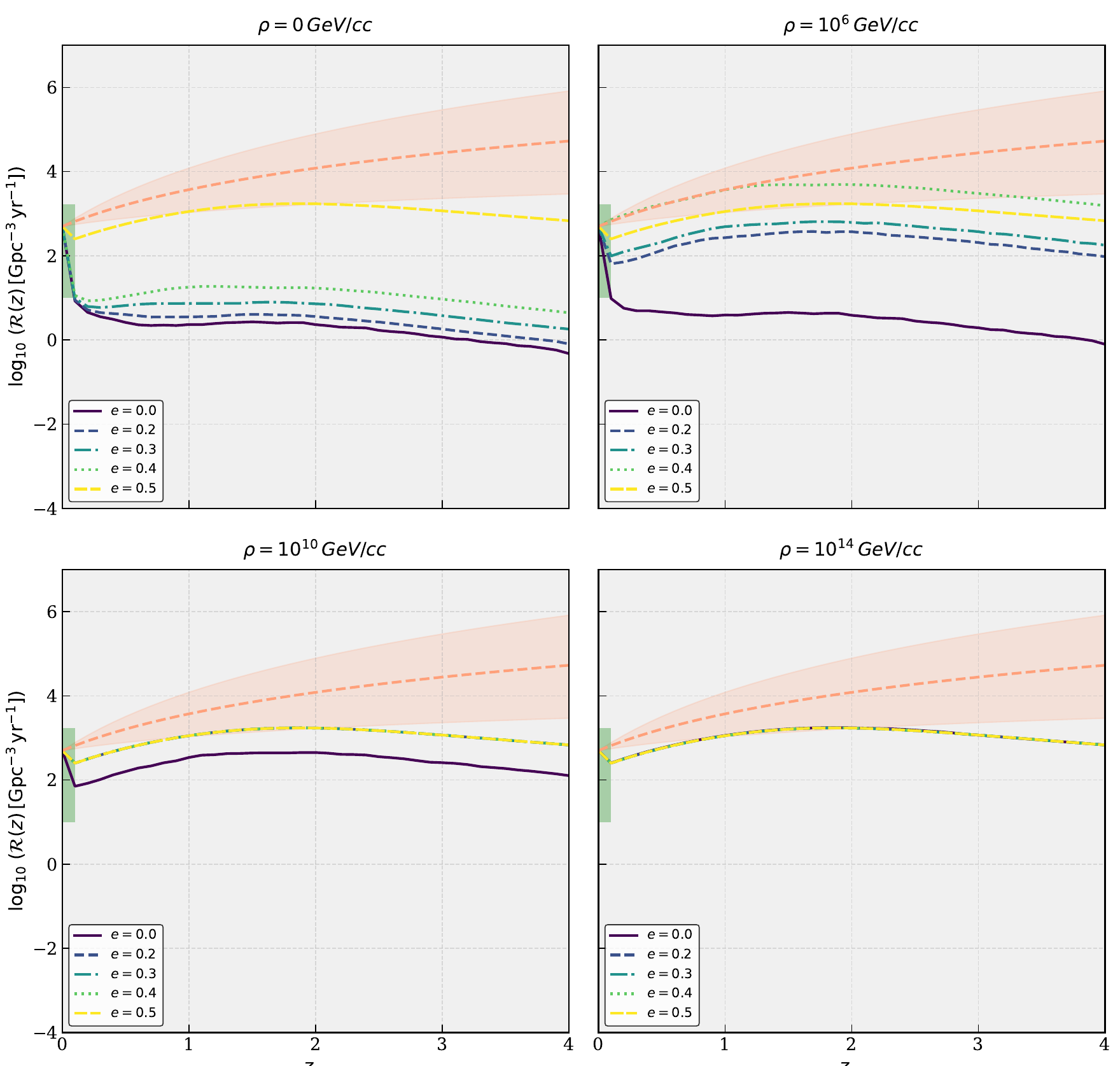}
    \caption{Merger Rate Density for Kroupa IMF, see caption of \ref{fig:flat_imf} for details.}
    \label{fig:kroupa_imf}
\end{figure*}

Then in order to make predictions for mergers at higher $z_m$ we have to extrapolate the trends that are observed within the sensitivity horizon of the respective detectors. This is usually done by a inferring a power-law behaviour given by a parameter $\kappa$ for higher $z_{\rm m}$ from GWTC-3 data. It also should be noted that the bounds on $\kappa$ come from the observations of BBH systems alone. This is simply because only BBH systems can be observed in enough numbers at $z_{\rm m}\lesssim1$ to make inferences about $\kappa$ using the GWTC-3 detector sensitivity. The overall $z_{\rm m}$-dependent statistics of merging compact binaries can then be modelled by a function with two parameters $R_0$ and $\kappa$. We note that the values of these parameters are based on direct observation and extrapolations from inferences respectively. Proceeding this way, we finally arrive at the inferred merger rate $\mathcal{R}_{\rm m}^\mathrm{inf}(z_{\rm m})$ given by,
\begin{equation}
\mathcal{R}_{\rm m}^\mathrm{inf}(z_{\rm m}) \simeq R_0\left(1+z_{\rm m}\right)^{\kappa}~.
\end{equation}
The greatest uncertainty to $R_0$ (in units of $\mathrm{Gpc}^{-3}\mathrm{yr}^{-1}$) is contributed by the binary NSs, yielding a very wide range for the merger rate $10 \leq R_{0}\leq 1700$ \cite{KAGRA:2021duu}. Consequently we have scaled our populations to have $R_{0}=500~\mathrm{Gpc}^{-3}\mathrm{yr}^{-1}$, so that our results are consistent with the bounds on $R_0$ presented by GWTC-3. 

Following the previous discussion, we have plotted our rescaled merger rate estimates and the median rate inferred from the GWTC-3 (orange dashed lines) for four different choices of the ULDM density. Then assuming the same allowed value of $R_{0}=500~\mathrm{Gpc}^{-3}\mathrm{yr}^{-1}$, we plot the uncertainty region mapped by the $90\%$ credible values of $\kappa$ as an orange shade (see Fig.~\ref{fig:flat_imf} and Fig.~\ref{fig:kroupa_imf}). In those figures the uncertainty in $R_{0}$ is also depicted as a green shaded region. Following the convention in \cite{Baibhav:2019gxm}, the extent of $R_{0}$ along the redshift is modelled till $z\sim0.1$, which is roughly the horizon size of the O3 LVK detectors. We also remark that the reconstruction of $\mathcal{R}_m^\mathrm{inf}(z_m)$ from GWTC-3 is mainly based on low redshift ($z_{\rm m}\lesssim 1$) events, and is therefore not likely to be an accurate representation at $z_{\rm m}\gtrsim 2$.

The rate estimates plotted for eccentric and circular cases in Fig.~\ref{fig:flat_imf} assuming a flat IMF and in Fig.~\ref{fig:kroupa_imf} assuming the Kroupa IMF constitute our second set of results. We note that in spite of the rescaling applied to $\mathcal{R}_{\rm m}(z_{\rm m})$, its $z_{\rm m}\lesssim 1$ behaviour for the most part does not respect the $90\%$ bounds on the power-law exponent as set by GWTC-3. However, for high enough ($\sim 0.5$) starting eccentricities and for $\rho_0 \sim \mathcal{O}(10^6) \text{GeV}/\textrm{cm}^{3}$ we see agreement of $\mathcal{R}(z)$ with $\mathcal{R}_ {\rm obs}(z)$ for $z\lesssim2$, irrespective of the choice of IMF. Interestingly we also observe that the agreement worsens as we move away from the optimal configuration as highlighted above. {Note that this criterion differs from the threshold $\rho_0\sim10^4$ GeV/cm$^3$ of Sec.~\ref{sec:merg_stat}, which marks where the ULDM significantly increases the merged fraction of eccentric binaries starting at $z=0.5$ (Fig.~\ref{fig:progmerg_ratio}), rather than where the agreement with the observed merger rate occurs.} We note from Eqs.~\eqref{eq:PDF_explicit} and Eq.~\eqref{eq:TDF_explicit} (see Sec. \ref{app:explicit}) that the dissipation terms involve a combination of $\rho_0$ with different non-trivial functions of orbital eccentricity, which are likely to make those terms optimum at an intermediate combination of $\rho_0$ and starting eccentricity. This indicates that the models that disagree with the bounds for all $z$ can be ruled out at $90\%$ confidence from the GWTC-3 data\footnote{Of course, there are several astrophysical uncertainties in our baseline model, starting from constant IMF, to ignoring Roche overflow.}. Note that we can rule out only the \textit{model}, which is a combination of our baseline model plus ULDM, given by its density $\rho_0$. We cannot however constrain ULDM $\rho_0$ in an absolute sense, because the effect of ULDM can be degenerate with the effects of additional dissipative factors like Roche overflow, stellar winds or CE phase evolution. Proceeding in this way, we see that for $\rho_0=10^{14} \text{GeV}/\textrm{cm}^{3}$, our model is only marginally consistent with the inferred merger rate for $z_{\rm m}\lesssim 2.0$ for all eccentricities and IMFs while $\rho_0 \lesssim 10^{6} \text{GeV}/\textrm{cm}^{3}, e \sim 0.5$ is the best performing. As expected, we also note that elliptic configurations predict a higher rate density than the quasi-circular ones when subjected to an identical set of underlying assumptions. This is along expected lines, given the smaller baseline merger timescales associated with eccentric binaries.

\begin{figure*}[!t]
    \centering
    \includegraphics[width=\textwidth]{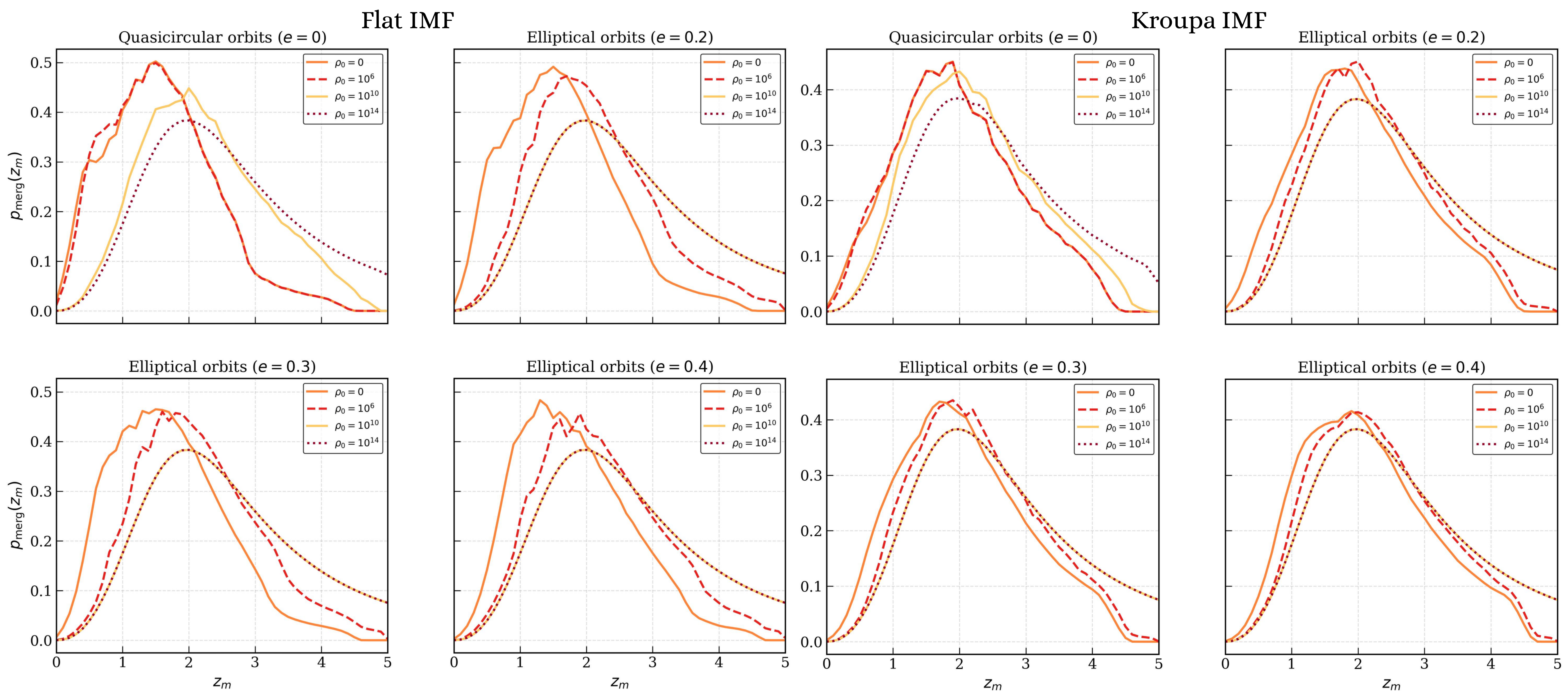}
    \caption{Plot showing the merger probability distribution $p_{\rm merg}(z_m)$ as a function of the merger redshift $z_m$ for four different choices of the ULDM density $\rho_0$. The left and right panels correspond to the Flat and Kroupa IMFs, respectively, while the different rows and columns show our choices of initial orbital eccentricity $e$. As evident, increasing $\rho_0$ generally shifts the merger distribution towards larger redshifts, although the magnitude of the effect depends on the initial eccentricity and the choice of IMF. See text for more details.}
    \label{Fig:Prob_flat_imf}
\end{figure*}

The merger rate density also decides the merger probability, which is simply the probability of having a binary merger event between the redshifts of $z_{\rm m}$ and $z_{\rm m}+dz_{\rm m}$, given by the time integral of the quantity $(d^2N_{\rm m}/dz_{\rm m}dt)$ defined above. In other words,
\begin{equation}
p_\mathrm{merg}(z_{\rm m}) \sim \int dt  \frac{d^{2}N_{\rm m}}{dz_{\rm m}dt}~.
\end{equation}
Figures \ref{Fig:Prob_flat_imf} show the merger-redshift probability distributions \(p_{\mathrm{merg}}(z_{\rm m})\) for the flat and Kroupa IMFs, respectively, considering both quasi-circular binaries and eccentric binaries with initial eccentricities \(e=0.2,\,0.3,\) and \(0.4\). We note that our findings are consistent with our previous results. Specifically, we obtained in Section~\ref{ssec:indbin} that denser ULDM environments tend to reduce merger timescales and make individual binaries marge faster, leading to a greater proportion of merged binaries among progenitors from a given source. In terms of a population with a source distribution, shorter merger timescales favour mergers at higher values of \(z_{\rm m}\). This trend is reflected in our results for \(p_{\mathrm{merg}}\) Fig. \ref{Fig:Prob_flat_imf}, although its effect on the peak position depends on the IMF and initial eccentricity. While the ULDM density $\rho_0=10^6 \text{GeV}/\textrm{cm}^{3}$ is indistinguishable from the baseline model without DM for quasi-circular binaries, its effects are more apparent for eccentric binaries. At higher densities, the distributions develop broader high-redshift tails, accompanied by a clearer shift of the peak towards higher \(z_{\rm m}\) for the flat IMF and a comparatively modest peak displacement for the Kroupa IMF. Moreover, for eccentric binaries, the distributions at \(\rho_0=10^{10}\) and \(10^{14}\,\mathrm{GeV/cm^{3}}\) nearly coincide. The redshift at the peak of the merger probability distribution is thus an observable, which carries the signature of ULDM density $\rho_0$, although they also depend on the IMF and initial eccentricity. However, with the current GW detector sensitivity, it is not possible to faithfully recreate the merger probability distribution of the GW events. We remark that improvements in ground-based GW detectors would overcome this difficulty and put another independent constraint upon the ULDM environments through direct observation of the peak in the merger redshift for merger probability distribution.

\section{Conclusion}\label{Sec:conc}

Given that astrophysical environments have previously been found to have non-trivial effects on the GW phasing and merger timescale of individual binaries, it is natural to expect that these environments also leave a significant imprint on their merger statistics. However, baryonic environments are highly case-specific and cannot be assumed to be present for every member of a merger population. In contrast, this situation is notably different for ULDM, as DM is expected to be present in all galaxies, and ULDM are also possibly present in the core region of a galaxy to avoid the core-cusp problem, where compact binary mergers can potentially occur. This ubiquity makes ULDM an intriguing candidate for studying environmental effects on merger statistics. Specifically, additional energy loss through dynamical friction, as the binaries moves in a DM environment, is expected to influence the distribution of the merger timescales, systematically shifting the merger populations toward shorter timescales. It is within this framework that the significance of our work becomes apparent.

In this study, we have performed a series of investigations aimed at quantifying the effects of ULDM and examining whether and how it can influence merger statistics of binaries. Our approach employs a simplified implementation of both the baseline astrophysical model and the DM medium. The models considered are however, just toy models, constructed with the primary objective of providing a minimal yet insightful framework to illustrate and analyze the influence of a dark matter environment on merger statistics. The goal is not to present a fully realistic astrophysical model but rather to explore the key qualitative effects that arise when compact binaries evolve within a ULDM background.

Despite these simplifying assumptions, our analysis clearly demonstrates that ULDM fields can indeed impact merger statistics, particularly when their density exceeds the threshold of $\rho_0>10^{2} \text{ GeV}/\textrm{cm}^{3}$ for elliptical case and $\rho_0>10^{8} \text{ GeV}/\textrm{cm}^{3}$ for quasicircular orbits. We started by considering individual binaries and then probing how their merger time gets modified. As expected, and as clearly depicted in Fig. \ref{fig:progmerg0p2}, the binaries in the DM environment merges quickly compared to the binary that evolves in vacuum. For example, in the case of a binary with total mass $45 M_{\odot}$ and mass-ratio of $0.8$, the binary with a ULDM density of $\rho_0=10^{8} \text{GeV}/\textrm{cm}^{3}$ merges, while the one without any DM environment does not merge, while they both start at $z=2.0$. Thus there is a significant effect on the merger timescale. 

This feature continues to hold even for a binary population, where also, the presence of ULDM leads to significantly higher proportion of mergers compared to the vacuum case. In particular, for $\rho_0\gtrsim \mathcal{O}(10^{8})\text{GeV}/\textrm{cm}^{3}$ for quasicircular orbits and $\rho_0\gtrsim \mathcal{O}(10^{4})\text{GeV}/\textrm{cm}^{3}$ for elliptical orbits, almost $60\%$ of the original progenitor binary configurations starting at $z = 0.5$ are merged. This also influences the merger rate of binaries. As evident from Fig. \ref{fig:flat_imf} and Fig. \ref{fig:kroupa_imf}, the DM densities below the threshold value of $10^{4} \text{GeV}/\textrm{cm}^{3}$, the merger rate is in tension with the LVK estimation of the merger rate. This seems to be a tantalizing evidence for the existence of DM, but then the uncertainties in the baseline model, along with various astrophysical considerations can tame this result. Thus, we would like to conclude that merger rate densities can provide a gold mine of information about the existence, and possibly, the nature of DM, provided we have better estimates of the merger rate at higher redshifts and more refined baseline model.  
Specifically, we have shown that increasing the density of the DM medium leads to a shift in the peak of the merger probability distribution (c.f. Fig. \ref{Fig:Prob_flat_imf}) towards higher redshifts. This is expected, as the DM density increases the dynamical friction also increases, leading to an enhanced loss of the orbital energy for the dipole. As a consequence, most of the binaries lose energy faster and hence merges at a higher redshift, i.e., earlier than expected from a vacuum environment. Thus the peak of the merger probability distribution in redshift is another observable that can tell us a lot about the existence and nature of the DM environment in galaxies. These results highlight the potential role of ULDM in shaping the observable characteristics of compact binary mergers and hence provides a novel perspective on how DM may influence astrophysical merger events.
We also note that, in localized ultralight-dark-matter configurations, the backreaction on the binary can acquire a nontrivial dependence on the orbital frequency. In particular, the periodic gravitational perturbation generated by the binary may excite wave-like modes of the ULDM distribution, leading to an enhancement of the effective dynamical friction in specific frequency ranges \cite{Tomaselli:2024bdd, Duque:2023seg}. Such frequency-dependent effects are not included in our present treatment, where the environmental backreaction is modeled through the local density and orbital velocity. Moreover our analysis also does not take into account the non-zero recoil of the centre of mass of the binary during drag and accretion as was identified in \cite{Cardoso:2020lxx} \footnote{We thank the anonymous referee for bringing these issues to our notice}. We leave such interesting questions as probable future extension of our work. 

\section*{Acknowledgements}

The authors acknowledge the use of the \textit{NOETHER} workstation at the Department of Physics, IIT Gandhinagar for computations. The authors acknowledge National Supercomputing Mission (NSM) for providing computing resources of ‘PARAM Ananta’ at IIT Gandhinagar, which is implemented by C-DAC and supported by the Ministry of Electronics and Information Technology (MeitY) and Department of Science and Technology (DST), Government of India. The research of KC is supported by the Ramanujan Fellowship (RJF/2025/000175) scheme
of the ANRF, Government of India. KC also acknowledges previous support from the PPLZ grant (Project No. 10005320/0501) of the Czech Academy of Sciences. The research of SS is supported by the Department of Science and Technology, Government of India, under the ANRF CRG Grant (No. CRG/2023/000545). The research of SC is supported by MATRICS (MTR/2023/000049) and Core Research (CRG/2023/000934) Grants from SERB, ANRF, Government of India. SC also thanks the local hospitality at ICTS through the associateship program, where a part of this work was done.

\clearpage
\onecolumngrid

\appendix

\section{Secular evolution of eccentric binaries}
\label{app:eccentric}

In this appendix we derive the evolution equations for the semi-major axis $a$ and the eccentricity $e$ of a binary moving through the ULDM medium, generalising the quasi-circular equation~(10) to eccentric orbits. Throughout, $M_1$ and $M_2$ denote the component masses, $M_T=M_1+M_2$ the total mass and $m_r=M_1M_2/M_T$ the reduced mass, and we use $G=c=1$.

\subsection{Balance laws and orbital averaging}
\label{app:balance}

At Newtonian order the orbital energy and the magnitude of the orbital angular momentum of the relative orbit are
\begin{equation}
E=-\frac{m_r M_T}{2a},
\qquad
L=m_r\sqrt{M_T\,a\,(1-e^2)} .
\label{eq:EL}
\end{equation}
The binary loses energy and angular momentum to gravitational radiation and to ULDM dynamical friction, so that
\begin{subequations}
\label{eq:balance}
\begin{align}
\frac{dE}{dt}&=-\left\langle P_{\rm GW}\right\rangle - \left\langle P_{\rm DF}\right\rangle,
\label{eq:Ebalance}\\
\frac{dL}{dt}&=-\left\langle T_{\rm GW}\right\rangle-\left\langle T_{\rm DF}\right\rangle ,
\label{eq:Lbalance}
\end{align}
\end{subequations}
where all fluxes are defined such that the we can consistently ensure the loss of angular momentum and energy

We work with secular, orbit-averaged equations. For any quantity $X$ evaluated on the Keplerian orbit,
\begin{equation}
\begin{aligned}
\langle X\rangle&=\frac{1}{T}\int_0^{T}X\,dt
=\frac{1}{2\pi}\int_0^{2\pi}X(\psi)\,\bigl(1-e\cos\psi\bigr)\,d\psi
\end{aligned}
\label{eq:average}
\end{equation}
where $T=2\pi/\Omega$ with $\Omega=\sqrt{M_T/a^3}$, $\psi$ is the eccentric anomaly, defined by $r=a(1-e\cos\psi)$. The second form, in which $dt=(1-e\cos\psi)\,d\psi/\Omega$, is the one used to evaluate the averages below. After averaging, all velocity dependence is expressed through $a$, $e$ and the complete elliptic integrals $K(e^2)$ and $E(e^2)$. \\

\textbf{\textit{Post-Newtonian treatment of the ULDM terms} :}
For the velocity-dependent ULDM terms, the exact expressions are first expanded in the non-relativistic regime and retained through 2PN order, i.e. through relative corrections of order \(v_i^4\). Schematically, the dynamical-friction kernel contains terms of the form
\[
v_i\left[1+\mathcal O(v_i^2)+\mathcal O(v_i^4)\right],
\]
while the accretion kernel takes the form
\[
v_i^{-1}\left[1+\mathcal O(v_i^2)+\mathcal O(v_i^4)\right].
\]
The individual terms in these expansions are then orbit averaged.

\subsection{Evolution of the semi-major axis}
\label{app:adot}

The gravitational-wave contribution to Eq.~\eqref{eq:balance} is the orbit-averaged Peters--Mathews luminosity~\cite{Peters:1963,Peters:1964},
\begin{equation}
\left\langle P_{\rm GW}\right\rangle
=\frac{32}{5}\,\frac{m_r^2M_T^3}{a^5(1-e^2)^{7/2}}
\left(1+\frac{73}{24}e^2+\frac{37}{96}e^4\right),
\label{eq:PGW}
\end{equation}
which reduces to \eqref{eq:pgw} for $e=0$. The dynamical-friction contribution follows from \eqref{eq:pdf}, summed over the two components,
\begin{equation}
\left\langle P_{\rm DF}\right\rangle
=\frac{4\pi n\hbar}{\mu}\sum_{i=1}^{2}\left\langle v_i\,\eta_i^2\,\Delta\Psi_i\right\rangle,
\label{eq:PDF}
\end{equation}
with, as in Sec.~II\,A,
\begin{equation}
\eta_i=-\frac{\gamma_i\,\mu M_i\,(1+v_i^2)}{v_i},
\qquad
\gamma_i=\frac{1}{\sqrt{1-v_i^2}},
\label{eq:eta}
\end{equation}
where $v_i$ is the speed of component $i$ relative to the ULDM rest frame and $\Delta\Psi_i=\mathrm{Re}\bigl[\Psi(1+\ell_{m}+i\eta_i)-\Psi(1+i\eta_i)\bigr]$ is the digamma difference as given in \eqref{eq:pdf}: it collects the contribution of the scattered partial waves up to the angular cut-off $\ell_{m}$ set by the maximum impact parameter. The orbit average is taken after the 2PN expansion described above.

The masses change because of accretion at the rate of \eqref{eq:m_dot}; in the averaged equations each component enters through
\begin{equation}
\left\langle \dot M_i\right\rangle\equiv\left\langle \dot M_i\big|_{Eq.\eqref{eq:m_dot}}\right\rangle ,
\label{eq:Mdot}
\end{equation}
obtained with the same 2PN expansion and orbital average.

Differentiating $E$ in Eq.~\eqref{eq:EL} gives
\begin{equation}
\frac{dE}{dt}=\frac{m_rM_T}{2a^2}\,\dot a-\frac{\dot M_1M_2+M_1\dot M_2}{2a} .
\label{eq:dEdt}
\end{equation}
Equating this to the energy balance~\eqref{eq:Ebalance} and solving for $\dot a$ yields
\begin{equation}
\dot a=-\frac{2a^2}{m_rM_T}\Bigl[\left\langle P_{\rm GW}\right\rangle+\left\langle P_{\rm DF}\right\rangle+\left\langle P^{(a)}_{\rm acc}\right\rangle\Bigr],
\label{eq:adot}
\end{equation}
where the accretion terms have been collected into the effective power
\begin{equation}
\left\langle P^{(a)}_{\rm acc}\right\rangle\equiv-\frac{1}{2a}\Bigl(M_2\left\langle\dot M_1\right\rangle+M_1\left\langle\dot M_2\right\rangle\Bigr).
\label{eq:Pacca}
\end{equation}

The quasi-circular evolution also follows from Eq.~\eqref{eq:adot} by taking the \(e\to0\) limit of the Peters--Mathews luminosity in Eq.~\eqref{eq:PGW}, which recovers Eq.~\eqref{eq:pgw}. In this limit, the separation and component speeds are constant over an orbital period at fixed secular parameters, so the dynamical-friction and accretion terms require no explicit orbital averaging.

\subsection{Evolution of the eccentricity}
\label{app:edot}

The gravitational-wave torque is the orbit-averaged Peters result~\cite{Peters:1964},
\begin{equation}
\left\langle T_{\rm GW}\right\rangle
=\frac{32}{5}\,\frac{m_r^2M_T^{5/2}}{a^{7/2}(1-e^2)^{2}}\left(1+\frac{7}{8}e^2\right),
\label{eq:TGW}
\end{equation}
and the dynamical-friction torque is kept in the compact form $\langle T_{\rm DF}\rangle$ until the end of this appendix.

Writing $L^2=M_1^2M_2^2\,a\,(1-e^2)/M_T$ and differentiating logarithmically,
\begin{equation}
\frac{\dot L}{L}=\sum_{i=1}^{2}\dot M_i\left(\frac{1}{M_i}-\frac{1}{2M_T}\right)+\frac{\dot a}{2a}-\frac{e\,\dot e}{1-e^2}.
\label{eq:dLdt}
\end{equation}
We impose the angular-momentum balance~\eqref{eq:Lbalance}, $$\dot L=-\langle T_{\rm GW}\rangle-\langle T_{\rm DF}\rangle$$ and eliminate $\dot a$ with Eqs.~\eqref{eq:adot} and \eqref{eq:Pacca}.

The accretion terms combine as $\sum_i\langle\dot M_i\rangle\bigl(3M_T-M_i\bigr)/(2M_iM_T)$, and solving for $\dot e$ gives
\begin{equation}
\begin{aligned}
\dot e={}&\frac{\sqrt{1-e^2}}{e\,m_r\sqrt{M_T a}}\Bigl[\left\langle T_{\rm GW}\right\rangle+\left\langle T_{\rm DF}\right\rangle\Bigr]\\
&-\frac{a(1-e^2)}{e\,m_rM_T}\Bigl[\left\langle P_{\rm GW}\right\rangle+\left\langle P_{\rm DF}\right\rangle+\left\langle P^{(e)}_{\rm acc}\right\rangle\Bigr],
\end{aligned}
\label{eq:edot}
\end{equation}
with
\begin{equation}
\left\langle P^{(e)}_{\rm acc}\right\rangle\equiv-\frac{m_r}{2a}\sum_{i=1}^{2}\frac{3M_T-M_i}{M_i}\left\langle\dot M_i\right\rangle .
\label{eq:Pacce}
\end{equation}
Note that the accretion terms enter the $a$ and $e$ equations through different effective powers, $\langle P^{(a)}_{\rm acc}\rangle\neq\langle P^{(e)}_{\rm acc}\rangle$, since the masses appear differently in $E$ and in $L$. As a check, inserting Eqs.~\eqref{eq:PGW} and \eqref{eq:TGW} alone into Eq.~\eqref{eq:edot} reproduces the Peters result \cite{Peters:1963, Peters:1964}.

\subsection{Coupled secular equations}
\label{app:final}

Collecting Eqs.~\eqref{eq:adot} and \eqref{eq:edot}, the orbit-averaged evolution of an eccentric binary in the ULDM medium is governed by
\begin{equation}
\boxed{
\begin{aligned}
\dot a={}&-\frac{2a^2}{m_rM_T}\Bigl[\left\langle P_{\rm GW}\right\rangle+\left\langle P_{\rm DF}\right\rangle+\left\langle P^{(a)}_{\rm acc}\right\rangle\Bigr],\\[6pt]
\dot e={}&\frac{\sqrt{1-e^2}}{e\,m_r\sqrt{M_T a}}\Bigl[\left\langle T_{\rm GW}\right\rangle+\left\langle T_{\rm DF}\right\rangle\Bigr]\\
&-\frac{a(1-e^2)}{e\,m_rM_T}\Bigl[\left\langle P_{\rm GW}\right\rangle+\left\langle P_{\rm DF}\right\rangle+\left\langle P^{(e)}_{\rm acc}\right\rangle\Bigr],
\end{aligned}}
\label{eq:coupled}
\end{equation}
with $\langle P_{\rm GW}\rangle$ and $\langle T_{\rm GW}\rangle$ given by Eqs.~\eqref{eq:PGW} and \eqref{eq:TGW}. The system~\eqref{eq:coupled} is integrated until the pericentre distance $a(1-e)$ reaches the collision radius of the pair (merger), and once $e\le 10^{-3}$ the evolution is continued with the quasi-circular equation~(10).

\subsection{Explicit orbit-averaged dissipation terms}
\label{app:explicit}

We finally list the closed-form averages entering Eq.~\eqref{eq:coupled}. They are expressed through the complete elliptic integrals $K\equiv K(e^2)$ and $E\equiv E(e^2)$ of the first and second kind and the dimensionless parameters
\begin{equation}
\alpha_i=\frac{M_i^2}{a\,M_T},
\qquad
\mathcal{M}_i=\mu M_i,
\qquad
\Lambda_i=\ell_{m,i}\,(\ell_{m,i}+1),
\label{eq:aux1}
\end{equation}

\textbf{\textit{Dynamical-friction power} : }
Using $\Delta\Psi_i=\sum_{k=1}^{\ell_{m,i}}k/(k^2+\eta_i^2)$, expanding to 2PN order and averaging,

\begin{equation}
\left\langle P_{\rm DF}\right\rangle
=
4\pi\rho_0
\sum_{i=1}^{2}
m_i^2
\frac{\Lambda_i\sqrt{\alpha_i}}
{6\pi\mathcal{M}_i^4}
\left[
\mathcal{E}_i\,E(e^2)
-
\mathcal{K}_i\,K(e^2)
\right].
\label{eq:PDF_explicit}
\end{equation}

with
\begin{align}
\mathcal{E}_i
={}&
\frac{1}{1- e^2}
\Big[
6(1 - e^2)\mathcal{M}_i^4
+3(1 - e^2)\Lambda_i\mathcal{M}_i^2\alpha_i
\nonumber\\
&\qquad
+(9 - e^2)\Lambda_i
\left(-1+2\Lambda_i+9\mathcal{M}_i^2\right)
\alpha_i^2
\Big],
\\[4pt]
\mathcal{K}_i
={}&
2\Lambda_i\alpha_i
\Big[
4(-1+2\Lambda_i)\alpha_i
+\mathcal{M}_i^2(3+36\alpha_i)
\Big].
\end{align}

\textbf{\textit{Accretion powers} :}
In the regime $|\eta_i|\gg1$ the 2PN-expanded accretion rate of Eq.~(5) averages to
\begin{equation}
\left\langle\dot M_i\right\rangle=\rho_0\,A_i\,\mathcal{R}_i\,\mathcal{F}_i,
\qquad
\mathcal{R}_i=\frac{1-R_i}{1+R_i},
\label{eq:Mdot_explicit}
\end{equation}
with
\begin{equation}
\mathcal{F}_i=\frac{4\mathcal{M}_i}{\sqrt{\alpha_i}}\Bigl[2\bigl(1+2\alpha_i^2\bigr)K-\bigl(1-2\alpha_i+2\alpha_i^2\bigr)E\Bigr],
\label{eq:F_explicit}
\end{equation}

where $A_i$ is the horizon area and $R_i$ the reflectivity of component $i$ (Eq.~(5)), so that $\mathcal{R}_i=1$ for a black hole and $\mathcal{R}_i=0$ for a star. Inserting Eq.~\eqref{eq:Mdot_explicit} into Eqs.~\eqref{eq:Pacca} and \eqref{eq:Pacce},
\begin{align}
\left\langle P^{(a)}_{\rm acc}\right\rangle&=-\frac{\rho_0}{2a}\Bigl(M_2A_1\mathcal{R}_1\mathcal{F}_1+M_1A_2\mathcal{R}_2\mathcal{F}_2\Bigr),
\label{eq:Pacca_explicit}\\
\left\langle P^{(e)}_{\rm acc}\right\rangle&=-\frac{m_r\,\rho_0}{2a}\sum_{i=1}^{2}\frac{3M_T-M_i}{M_i}\,A_i\mathcal{R}_i\mathcal{F}_i .
\label{eq:Pacce_explicit}
\end{align}

\textbf{\textit{Dynamical-friction torque} : }
For the torque we use the leading-order dynamical-friction torque of Ref.~\cite{Roy:2024rhe}, where $v$ is the relative orbital speed and $Q=q/(1+q)^2$ the symmetric mass ratio with $q=M_2/M_1\le1$. Its orbit average Eq.~\eqref{eq:average}, is
\begin{equation}
\left\langle T_{\rm DF}\right\rangle=\frac{4\pi\rho_0a^2M_T(1-2Q)}{Q^2}\,
\frac{5(1-e^2)^{3/2}+12e^2-4}{(1-e^2)^{3/2}},
\label{eq:TDF_explicit}
\end{equation}
\section{Change in orbital parameters across a stellar endpoint}
\label{app:eccentric_endpoint}

For an eccentric orbit, the orbital angular momentum may be written as
\begin{equation}
L=m_r r^2\Omega ,
\end{equation}
where $\Omega=\dot{\phi}$. It is useful to introduce the specific
angular momentum
\begin{equation}
h\equiv \frac{L}{m_r}=r^2\Omega=r^2\dot{\phi}.
\label{eq:specific_ang_mom}
\end{equation}
We first obtain the corresponding form of Kepler's third law for an
elliptic orbit. The orbital period is
\begin{equation}
T
=
\int_0^T dt
=
\int_0^{2\pi}\frac{d\phi}{\dot{\phi}}.
\label{eq:period_phi}
\end{equation}
Using Kepler's law of areas,
\begin{equation}
\frac{dA}{dt}
=
\frac{1}{2}r^2\dot{\phi}
=
\frac{h}{2},
\end{equation}
and hence, over one complete orbit,
\begin{equation}
A
=
\int_0^T \frac{dA}{dt}\,dt
=
\frac{hT}{2}.
\end{equation}
Since the area of an ellipse is $A=\pi ab$, it follows that
\begin{equation}
2\pi ab=hT,
\end{equation}
or
\begin{equation}
T=\frac{2\pi ab}{h}.
\label{eq:period_area}
\end{equation}
For an elliptic Keplerian orbit,
\begin{equation}
h^2=M_Ta(1-e^2),
\qquad
b^2=a^2(1-e^2).
\end{equation}
Therefore,
\begin{align}
T^2
&=
\frac{4\pi^2a^2b^2}{h^2}
\nonumber\\
&=
\frac{
4\pi^2a^4(1-e^2)
}{
aM_T(1-e^2)
}
\nonumber\\
&=
\frac{4\pi^2a^3}{M_T}.
\label{eq:kepler_eccentric}
\end{align}
We define
\begin{equation}
\mathcal I_\Omega
\equiv
\int_0^{2\pi}\frac{d\phi}{\Omega}.
\label{eq:Iomega_def}
\end{equation}
Thus $\mathcal I_\Omega=T$, and Eq.~\eqref{eq:kepler_eccentric} gives
\begin{equation}
\mathcal I_\Omega^2
=
\frac{4\pi^2a^3}{M_T}.
\label{eq:Iomega_kepler}
\end{equation}
Taking the fractional variation of this relation gives
\begin{equation}
3\frac{\delta a}{a}
-
2\frac{\delta\mathcal I_\Omega}{\mathcal I_\Omega}
=
\frac{\delta M_T}{M_T}.
\label{eq:delta_kepler_ecc}
\end{equation}
We now consider the corresponding variation of the orbital angular
momentum. From Eq.~\eqref{eq:specific_ang_mom},
\begin{equation}
r^2=\frac{h}{\Omega},
\end{equation}
and hence the corresponding averaged quantities satisfy
\begin{equation}
m_r\left\langle r^2\right\rangle
=
L\,\mathcal I_\Omega.
\label{eq:r2_relation}
\end{equation}
Taking the fractional variation gives
\begin{equation}
\frac{\delta m_r}{m_r}
+
\frac{\delta\left\langle r^2\right\rangle}
     {\left\langle r^2\right\rangle}
=
\frac{\delta L}{L}
+
\frac{\delta\mathcal I_\Omega}{\mathcal I_\Omega},
\end{equation}
or equivalently,
\begin{equation}
\frac{\delta L}{L}
=
\frac{\delta m_r}{m_r}
+
\frac{\delta\left\langle r^2\right\rangle}
     {\left\langle r^2\right\rangle}
-
\frac{\delta\mathcal I_\Omega}{\mathcal I_\Omega}.
\label{eq:deltaL_general}
\end{equation}
For an eccentric orbit,
\begin{equation}
\left\langle r^2\right\rangle
=
a^2
\left(
1+\frac{3}{2}e^2
\right),
\label{eq:r2_average}
\end{equation}
so that
\begin{equation}
\frac{
\delta\left\langle r^2\right\rangle
}{
\left\langle r^2\right\rangle
}
=
2\frac{\delta a}{a}
+
\frac{
\delta\left(1+\frac{3}{2}e^2\right)
}{
1+\frac{3}{2}e^2
}.
\label{eq:delta_r2}
\end{equation}
Substituting Eq.~\eqref{eq:delta_r2} into
Eq.~\eqref{eq:deltaL_general}, we obtain
\begin{equation}
\frac{\delta L}{L}
=
\frac{\delta m_r}{m_r}
+
2\frac{\delta a}{a}
+
\frac{
\delta\left(1+\frac{3}{2}e^2\right)
}{
1+\frac{3}{2}e^2
}
-
\frac{\delta\mathcal I_\Omega}{\mathcal I_\Omega}.
\label{eq:deltaL_eccentric}
\end{equation}
Together with
\begin{equation}
3\frac{\delta a}{a}
-
2\frac{\delta\mathcal I_\Omega}{\mathcal I_\Omega}
=
\frac{\delta M_T}{M_T},
\end{equation}
we may eliminate $\delta a/a$. Multiplying
Eq.~\eqref{eq:deltaL_eccentric} by three and using
Eq.~\eqref{eq:delta_kepler_ecc} gives
\begin{equation}
3\frac{\delta L}{L}
=
2\frac{\delta M_T}{M_T}
+
3\frac{\delta m_r}{m_r}
+
3
\frac{
\delta\left(1+\frac{3}{2}e^2\right)
}{
1+\frac{3}{2}e^2
}
+
\frac{\delta\mathcal I_\Omega}{\mathcal I_\Omega}.
\label{eq:deltaL_eliminated}
\end{equation}
Equivalently,
\begin{equation}
-\frac{\delta\mathcal I_\Omega}{\mathcal I_\Omega}
=
2\frac{\delta M_T}{M_T}
+
3\frac{\delta m_r}{m_r}
-
3\frac{\delta L}{L}
+
3
\frac{
\delta\left(1+\frac{3}{2}e^2\right)
}{
1+\frac{3}{2}e^2
}.
\label{eq:Iomega_variation}
\end{equation}
On the other hand, Eq.~\eqref{eq:delta_kepler_ecc} may be written as
\begin{equation}
3\frac{\delta a}{a}
=
\frac{\delta M_T}{M_T}
+
2\frac{\delta\mathcal I_\Omega}{\mathcal I_\Omega}.
\label{eq:deltaa_eccentric}
\end{equation}
Now, we opt for a model which is minimally deviated from the corresponding quasi-circular case, so that the eccentricity effect contributes solely during evolution and not during mass-transfer. From Eq.~\eqref{eq:deltaa_eccentric}, we see that $\delta a/a \leq (1/3)$ only if the eccentricity term in $\delta{\mathcal I_\Omega}/{\mathcal I_\Omega}\leq 0$
For the minimal eccentric model adopted here, this is most straightforwardly implemented take the change in
eccentricity across the stellar endpoint to vanish (c.f Eq.~\eqref{eq:Iomega_variation}). Hence,
\begin{equation}
3
\frac{
\delta\left(1+\frac{3}{2}e^2\right)
}{
1+\frac{3}{2}e^2
}
=
3\,
\frac{3e\,\delta e}
{1+\frac{3}{2}e^2}
=0,
\end{equation}
which gives
\begin{equation}
\boxed{\delta e=0}.
\label{eq:deltae_endpoint}
\end{equation}
The change in the semi-major axis is chosen analogously to the
quasi-circular prescription,
\begin{equation}
\boxed{
\frac{\delta a}{a}
=
\frac{\delta M_T}{M_T}
}.
\label{eq:deltaa_endpoint}
\end{equation}
Thus, across a stellar endpoint, the prescription used for an eccentric
binary is the same as that of a quasicircular case
\begin{equation}
\boxed{
\frac{\delta a}{a}
=
\frac{\delta M_T}{M_T},
\qquad
\delta e=0.
}
\label{eq:endpoint_eccentric_final}
\end{equation}

\twocolumngrid
\bibliographystyle{IEEEtran}  
\bibliography{ref}

\end{document}